\documentstyle[preprint,aps,prb]{revtex}
\begin{document}
\title{Dynamics of quantum spin systems \\ in dimer and valence-bond-solid
ground states \\ stabilized by competing interactions}
\author{Yongmin Yu and Gerhard M\"uller}
\address{Department of Physics, The University of Rhode Island, Kingston,
Rhode Island 02881-0817}
\date{\today}
\maketitle
\begin{abstract}
For special coupling ratios, the one-dimensional (1D) $s=1/2$ Heisenberg
model with antiferromagnetic nearest and next-nearest neighbor interactions
has a pure dimer ground state, and the 1D $s=1$ Heisenberg model with
antiferromagnetic bilinear and biquadratic interactions has an exact
valence-bond-solid ground state. The recursion method is used to calculate
the $T=0$ spin dynamic structure factor for both models and, for the
$s=1/2$ model, also the dimer dynamic structure factor.
New results for line shapes and dynamically relevant
dispersions are obtained.
\end{abstract}

%\twocolumn

Correlated quantum fluctuations in the ground state are a generic feature
of quantum many-body systems.
They make it hard to take finite-size effects into account in computational
studies of zero-temperature dynamical properties.
Interestingly, there are several known cases, where a relatively
simple-structured ground state is stabilized by competing terms in the
microscopic Hamiltonian.
This ground state may or may not be long-range ordered.
The essential attribute is that its fluctuations are not correlated or
only over a short distance on the lattice.
This phenomenon typically occurs with no accompanying
simplification in the excitation spectrum or any dynamical quantity.
Nevertheless, any such situation provides an unsuspected window for
dynamical studies which promise to be much less plagued with finite-size
effects than is typically the case.
The recursion method\cite{rm} in combination with recently developed
techniques of continued-fraction analysis\cite{VM94} is an ideal
calculational tool for that purpose, the key property being that it
extracts the dynamical information from the ground-state wave function.

In a previous paper we have reported the study of one such case, namely
the one-dimensional (1D) spin-$s$ $XYZ$ model.\cite{VSM94}
In a magnetic field of particular strength, this model has a product
ground state with spontaneous ferro- or antiferromagnetic long-range
order perpendicular to the field.
Here we present new results for two different models with simple-structured
ground states.

The first model is the 1D $s=1/2$ Heisenberg antiferromagnet with
competing nearest and next-nearest neighbor interactions,
\begin{equation}
H=\sum\limits_{l=1}^N\{J_1{\bf S}_l\cdot {\bf S}_{l+1}+
J_2{\bf S}_l\cdot {\bf S}_{l+2}\},
\label{1}
\end{equation}
with an even number of spins and periodic boundary conditions.
This system undergoes a $T=0$ phase transition at $J_2/J_1 \simeq 0.25$
from a spin-fluid phase to a phase with spontaneous dimer long-range
order.\cite{dim1}
In the (critical) spin-fluid phase, the correlations of the quantum
fluctuations are particularly strong.
In the dimer phase, their continued presence, albeit much attenuated,
manifests itself, for example, in the finite-size splitting of the
ground-state doublet.
The exception is the special coupling ratio $J_2/J_1=0.5$, where the pure
dimer ground-state is realized.\cite{dim2,SS81}
Here the ground-state energy per site is size-independent:
$E_0/N=-\frac 38J_1$.

The two (translationally invariant) dimer ground-state wave functions can
be expressed in terms of products of singlets formed by pairs of
nearest neighbor spins:
\begin{equation}
|\Phi_{\pm}\rangle =
\left(2\pm (-1/2)^{\frac{N-4}2}\right)^{-1/2}
\{|\Phi_1\rangle \pm |\Phi_2\rangle\},
\label{2}
\end{equation}
where $|\Phi_1\rangle = [1,2][3,4]\cdots[N-1,N]$, $|\Phi_2\rangle$ =
$[2,3][4,5]\cdots$ $[N,1]$,
$[l,l+1]=\{|\uparrow \downarrow \rangle -
|\downarrow \uparrow \rangle\}/\sqrt{2}$.

The dimer order parameter,
$D=N^{-1}\Sigma_l(-1)^lD_l$, $D_l$ =
$S_l^{+}S_{l+1}^{-}$ + $S_l^{-}S_{l+1}^{+}$,
has a nonzero expectation value, $\langle D \rangle = \pm 1/2$ in the
(non-orthogonal) symmetry-breaking states
$|\Phi_1\rangle$ and $|\Phi_2\rangle$.
The order-parameter correlation function in this case is
not a two-spin correlation function,
($\langle S_l^zS_{l+n}^z\rangle =0$ for $|n|>1$),
but a four-spin correlation function:
$\langle D_lD_{l+n}\rangle = (-1)^n/4$ for $n \neq 0$.
Hence it will be instructive to compare the spin dynamic structure factor
$S_{zz}(q,\omega)$ and the dimer dynamic structure factor
$S_{DD}(q,\omega)$, i.e. the function
\begin{equation}
S_{AA}(q,\omega )=\int\limits_{-\infty }^{+\infty }dt e^{i\omega
t}\langle A_q(t)A_{-q}\rangle \;,
\label{3}
\end{equation}
where $A_q$ stands for the spin fluctuation operator,
$S_q^z$ = $N^{-1/2}\Sigma_le^{iql}S_l^z$ or the dimer fluctuation operator,
$D_q$ = $N^{-1/2}\Sigma_le^{iql}[D_l-\langle D_l\rangle]$.

By means of the recursion method\cite{rm} in combination with
a strong-coupling continued-fraction analysis,\cite{VM94,VZSM94}
we calculate the dynamically relevant excitation spectra and
the spectral-weight distributions of these two functions.
The recursion algorithm in the present context is based on an orthogonal
expansion of the wave function $|\Psi_q^A(t)\rangle = A_q(-t)|\Phi\rangle$.
It produces (after some intermediate steps) a sequence of
continued-fraction coefficients $\Delta^A_1(q), \Delta^A_2(q), \ldots $
for the relaxation function,
\begin{equation}
c_0^{AA}(q,z) = \frac{1}{\displaystyle z + \frac{\Delta^{A}_1(q)}
{\displaystyle
z +  \frac{\Delta^{A}_2(q)}{\displaystyle z + \ldots }  }  }\;,
\label{4}
\end{equation}
which is the Laplace transform of the symmetrized correlation function
$\Re\langle A_q(t)A_{-q}\rangle/$ $\langle A_qA_{-q}\rangle$.
The $T=0$ dynamic structure factor (\ref{3}) is then obtained from
(\ref{4}) via
\begin{equation}
S_{AA}(q,\omega) =  4\langle A_qA_{-q}\rangle\Theta(\omega)\lim
 \limits_{\varepsilon
\rightarrow 0} \Re [c_{0}^{AA} (q, \varepsilon - i\omega)] \;.
\label{5}
\end{equation}
The simple $N$-dependence of the dimer ground-state wave functions
(\ref{2}) offers the advantage that we can compute a significant number of
$N$-independent coefficients $\Delta_k^{A}(q)$.
These data are the input to the well-tested strong-coupling
continued-fraction reconstruction.\cite{VM94,VZSM94}

The results for the frequency-dependence of the dynamic structure factor
$S_{zz}(q,\omega )$ at the wave numbers realized for $N=18$ are displayed
in Fig. \ref{F1}.
The set of curves is perfectly compatible with a function
$S_{zz}(q,\omega)$ that varies smoothly in $q$ as well as in $\omega$.
For every $q$-value we observe a single peak in the frequency range of
interest.
This peak is very broad at $q$ near zero or $\pi$.
The width shrinks as $q$ approaches $\pi /2$ from either side.

At $q=\pi /2$ (not realized for $N=18$) the dynamically relevant
excitation spectrum reduces to a single mode.
The states $S_{\pi/2}^z|\Phi_{\pm}\rangle$ are, in fact, known to be exact
eigenstates of the system.\cite{CM82,term}
The dynamically relevant dispersion of $S_{zz}(q,\omega)$ is symmetric
about $q=\pi /2$, where it has a smooth maximum at frequency
$\omega /J_1=1.0$.
For $q=0$ and $q=\pi$ it has smooth minima at $\omega/J_1\simeq 0.5$.

On the basis of a variational calculation for the pure dimer state,
Shastry and Sutherland\cite{SS81} obtained an excitation spectrum for this
model which consists of a continuum of two-defect scattering states with a
lower boundary $\epsilon (q)=J_1(\frac 54-|\cos q|)$ and, for the
restricted range $0.36\pi <q<0.64\pi$ of wave numbers, a branch of defect
bound states, which emerges from the lower continuum boundary and
has a smooth maximum reaching up to $\omega /J_1=1$ at $q=\pi/2$.

In this context, our results suggest that for $q$ near zero or $\pi $,
the spectral weight of $S_{zz}(q,\omega)$ is distributed over a broad
frequency range of two-defect scattering states.
As $q$ approaches $\pi/2$ from either side, the spectral weight is shifted
gradually to the two-defect bound state.

Our result for the dynamic structure factor $S_{DD}(q,\omega)$ is plotted
in Fig. \ref{F2} for the same set of wave numbers.
The line shapes and peak positions resemble those shown in Fig. \ref{1} for
$S_{zz}(q,\omega)$, but there are some notable differences:
The spectral weight in $S_{DD}(q,\omega)$ is concentrated at
somewhat higher energies.
The intensity at $q=0$ is nonzero.
The shape of the dimer dispersion is different.
$S_{DD}(\pi/2,\omega )$ does not reduce to a single line.
In $S_{DD}(\pi,\omega)$ the spectral weight is shared between a continuum
and a $\delta $-peak at $\omega =0$.\cite{note1}
The latter contribution reflects the presence of dimer long-range order in
the ground state.

The Hamiltonian of the 1D $s=1$ model with isotropic bilinear and
biquadratic exchange, our second example, is most conveniently expressed
in the form
\begin{equation}
H_\gamma =J\sum\limits_{l=1}^N\{\cos\gamma {\bf S}_l\cdot{\bf S}_{l+1}
+\sin\gamma({\bf S}_l\cdot{\bf S}_{l+1})^2\}
\label{6}
\end{equation}
with a single parameter $-\pi <\gamma \leq \pi $.
More than a decade of research on this model has established a $T=0$
phase diagram consisting of the short-range ordered Haldane phase and three
phases with dimer, trimer, and ferromagnetic long-range order.\cite{note2}
In the Haldane phase, which includes the Heisenberg antiferromagnet
($\gamma =0$), the ground state is a non-degenerate singlet state ($S_T=0$)
separated by a gap from the threshold of the excitation spectrum.

At the parameter value $\gamma =\arctan (1/3)\simeq 18.4^{\circ}$ within
this phase, the ground-state wave function is exactly known.\cite{AKLT87}
It is a realization of the so-called valence-bond-solid (VBS) wave
function, which can be assembled from the same parts as the dimer
state (\ref{2}).
The spin 1 at each lattice site is expressed as a spin-1/2 pair in a
triplet state.
The singlet-pair forming valence bond involves one fictitious spin 1/2
from each of two neighboring lattice sites.
The VBS state can then be regarded as a chain of valence bonds linking
successive symmetrized spin-1/2 pairs, which are given by the $S_T^z=0$
vector of the triplet on each lattice site.

The static spin correlation function in the VBS state,\cite{AKLT87}
$\langle S_l^zS_{l+n}^z\rangle$ = $\frac 43(-1)^n3^{-|n|}$, $(n\neq 0)$,
reflects magnetic short-range order with a very short correlation length:
$\xi =1/\ln3\simeq 0.91$.
The static structure factor is then a non-singular function
with a smooth minimum at $q=0$ and a smooth maximum at $q=\pi$:
\begin{equation}
S^{zz}(q)=2(1-\cos q)/(5+3\cos q).
\label{11}
\end{equation}
The simple structure of the VBS state makes this quantity free of
finite-size effects for $q=2\pi l/N$, $l=0,\ldots,N-1$.
Again, this simplification does not extend to the excitation spectrum and
the dynamical properties.

In Fig. \ref{F3} we have plotted $S_{zz}(q,\omega )$ as obtained
from a strong-coupling continued-fraction analysis with coefficients
extracted from the $N=12$ VBS wave function.
At each value of $q$ the spectral weight of $S_{zz}(q,\omega )$ is found
to be concentrated in a single peak with symmetric line shape.
The peak frequency decreases monotonically with $q$.
The suggested gap value at $q=\pi$ is $\Delta E/J\simeq 0.66$.
The linewidth tends to be very small at $q$ near $\pi $, where the peak
frequency is lowest and the intensity highest.
It gains considerably in breadth at $q$ near $0$, where the peak frequency is
higher and the intensity much lower.

The monotonic $q$-dependence of the dynamically relevant dispersion for
$S_{zz}(q,\omega )$ in the VBS state ($\gamma \simeq 18.4^{\circ }$) is
markedly different from the corresponding quantity in the Heisenberg
antiferromagnet ($\gamma =0$), which belongs to the same phase.
In the Heisenberg case, the dispersion has a smooth maximum
at $q\simeq\pi/2$ and smooth minima of unequal height at $q=0$ and
$q=\pi$.\cite{hbafm}

This work was supported by NSF Grant DMR-93-12252 and by the NCSA at
Urbana-Champaign.
\pagebreak

\pagebreak

\begin{figure}
\caption[one]
{Dynamic structure factor $S_{zz}(q,\omega)$ vs $\omega$ for
$q=2\pi l/N$, $l=0,1,\ldots,N/2$ in the dimer ground state
$|\Phi_{+}\rangle$ of the Hamiltonian (\ref{1}) at $J_2=0.5$ and $J_1=1$
with $N=18$, obtained via strong-coupling continued-fraction
reconstruction based on the coefficients $\Delta _1,...,\Delta _9$
and a Gaussian terminator as explained in Refs.
\onlinecite{VM94,VZSM94}.}
\label{F1}
\end{figure}

\begin{figure}
\caption[two]
{Dynamic structure factor $S_{DD}(q,\omega )$ vs $\omega$ for
$q=2\pi l/N$, $l=0,1,\ldots,N/2$ in the dimer ground state
$|\Phi_{+}\rangle$ of the Hamiltonian (\ref{1}) at $J_2=0.5$ and
$J_1=1$ with $N=18$, obtained via strong-coupling continued-fraction
reconstruction based on the coefficients $\Delta _1,...,\Delta _9$
and a Gaussian terminator.}
\label{F2}
\end{figure}

\begin{figure}
\caption[three]
{Dynamic structure factor $S_{zz}(q,\omega)$ vs $\omega $ for
$q=2\pi l/N$, $l=0,1,\ldots,N/2$ in the VBS ground state of the model
system (\ref{6}) at $J=1$ and $\gamma=\arctan(1/3)$ with $N=12$, obtained
via a strong-coupling continued-fraction analysis based on the coefficients
$\Delta_1,...,\Delta_6$ and a Gaussian terminator}
\label{F3}
\end{figure}


\begin{references}
\bibitem{note1}
The $\delta$-peak can be separated from the continuum part of
$S_{DD}(\pi,\omega)$ as explained in Sec. 9-8 of Ref. \onlinecite{VM94}.
\end{references}
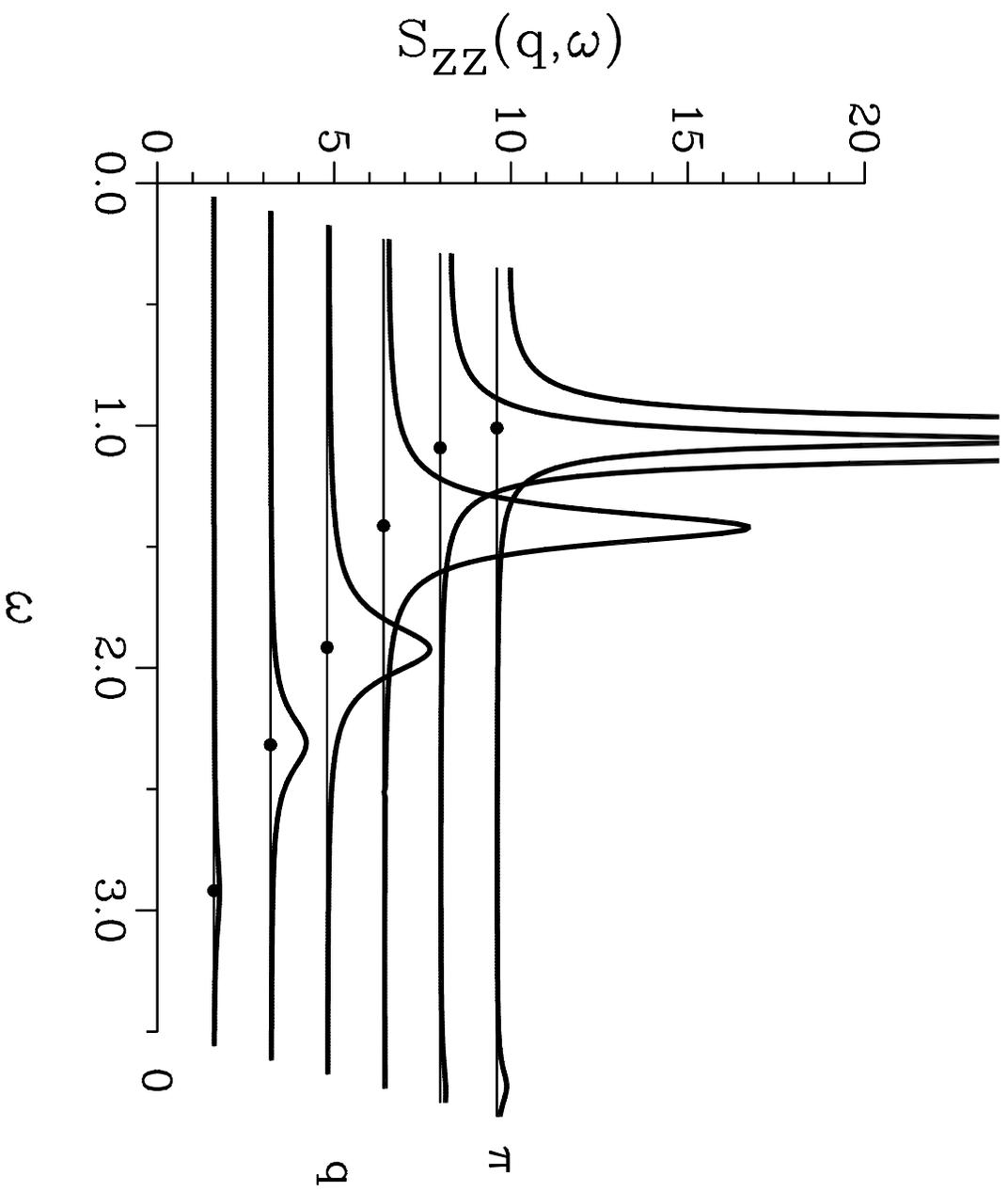
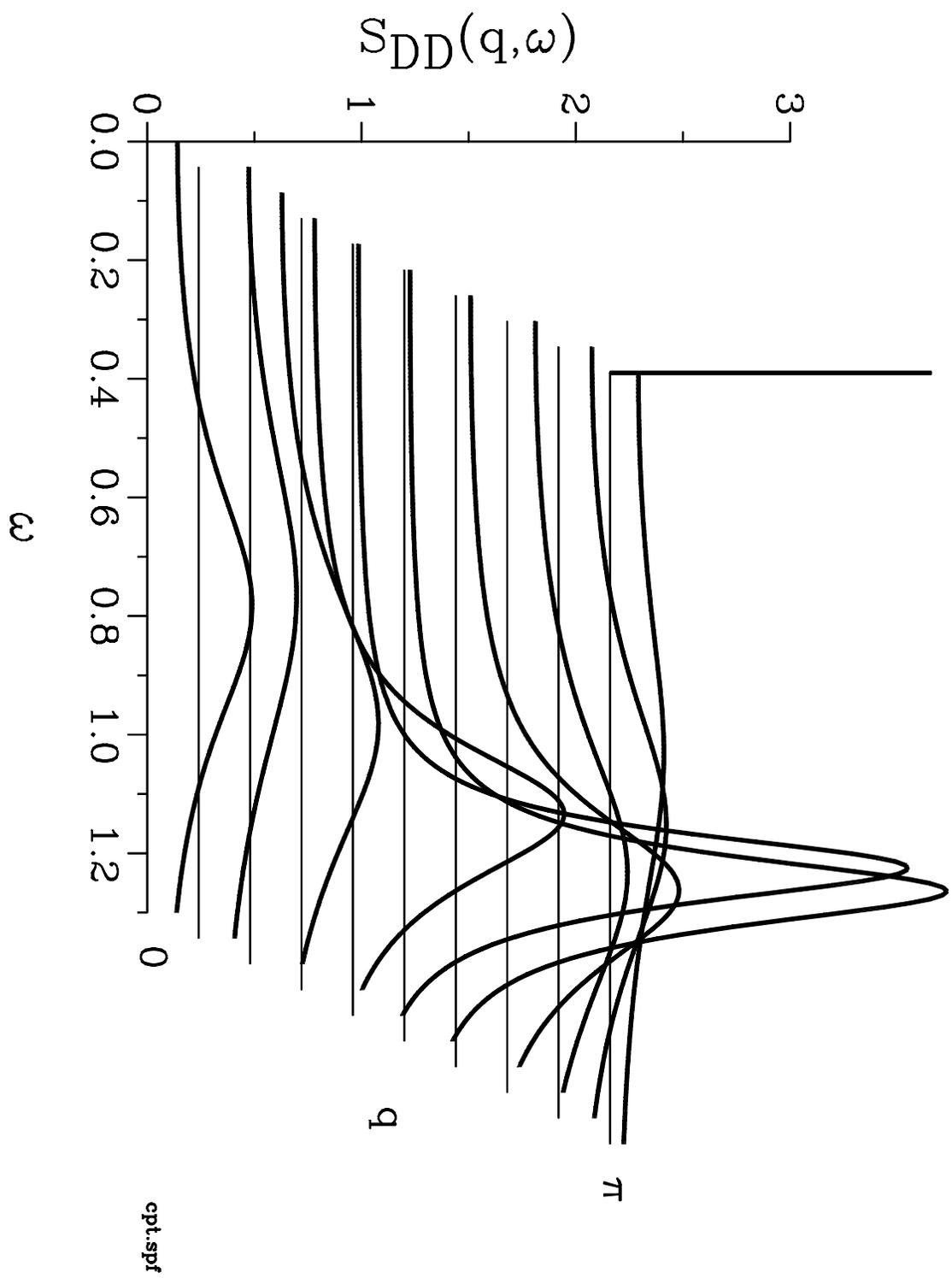
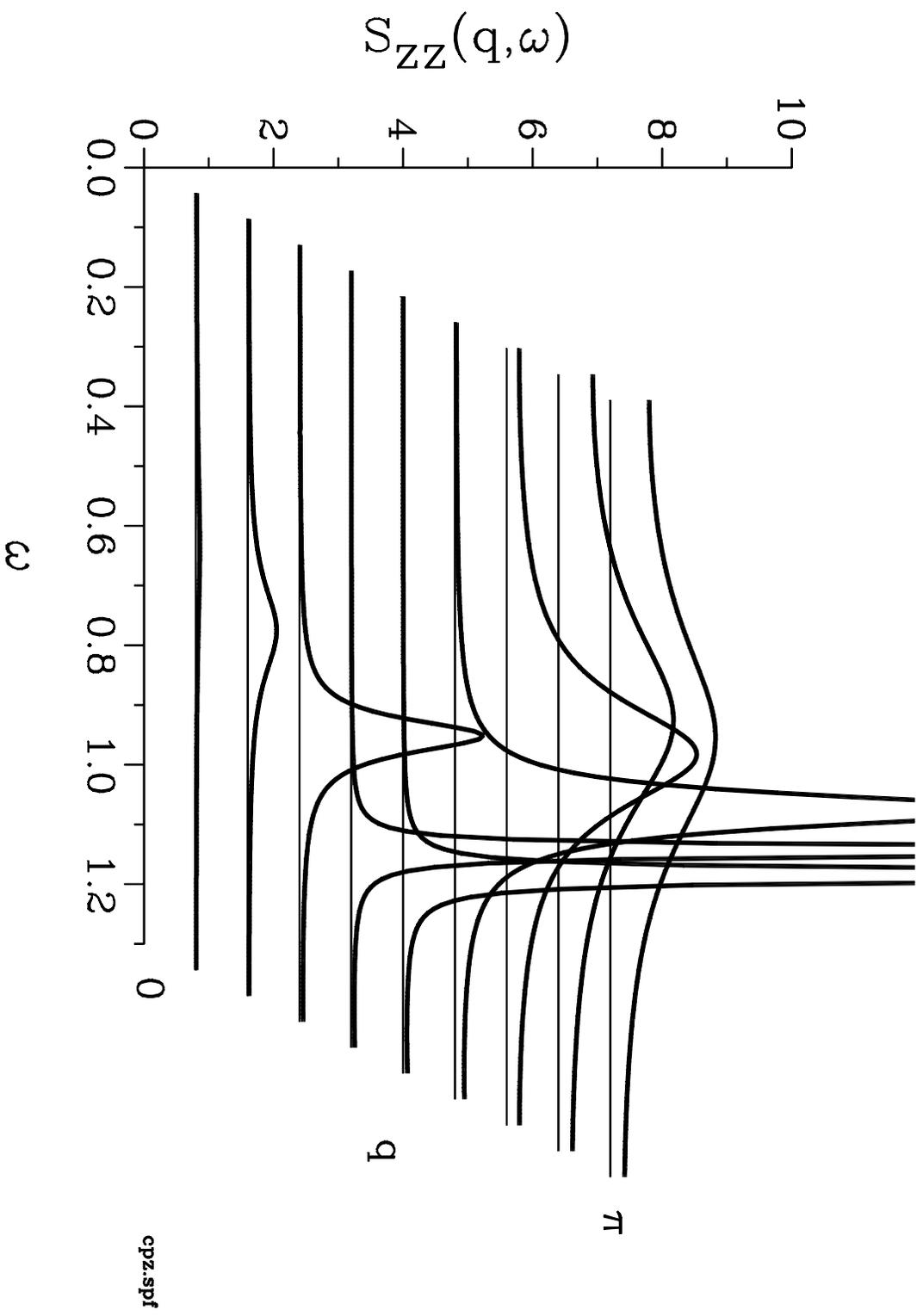
\end{document}